\documentclass[aps,prl,twocolumn,amsmath,amssymb,floatfix,bibnotes,preprintnumbers,superscriptaddress,longbibliography]{revtex4-2}
\usepackage[T1]{fontenc}
\usepackage[utf8]{inputenc}
\usepackage{amsmath}
\usepackage{graphicx,epstopdf}
\usepackage{gensymb}
\usepackage{float}
\usepackage{dcolumn,booktabs,microtype,afterpage}
\usepackage{amssymb}
\usepackage{sidecap}
\usepackage[mathlines]{lineno}
\makeatletter
\usepackage{color}
\usepackage[colorlinks,bookmarks=false,citecolor=darkblue,linkcolor=red,urlcolor=blue]{hyperref}

\definecolor{darkred}{rgb}{0.7,0.0,0.0}

\definecolor{darkblue}{rgb}{0,0.02,0.45}

\definecolor{darkgreen}{rgb}{0.02,0.45,0.0}

\definecolor{violet}{rgb}{0.8,0.2,0.6}

\DeclareUnicodeCharacter{2212}{-}

\begin{document}
	
\title{Contrasting $c$-axis and in-plane uniaxial stress effects on superconductivity and stripe order in La$_{1.885}$Ba$_{0.115}$CuO$_4$}
	
\author{S.S.~Islam}
\affiliation{PSI Center for Neutron and Muon Sciences CNM, 5232 Villigen PSI, Switzerland}
	
\author{V.~Sazgari}
\affiliation{PSI Center for Neutron and Muon Sciences CNM, 5232 Villigen PSI, Switzerland}

\author{J.N.~Graham}
\affiliation{PSI Center for Neutron and Muon Sciences CNM, 5232 Villigen PSI, Switzerland}

\author{O.~Gerguri}
\affiliation{PSI Center for Neutron and Muon Sciences CNM, 5232 Villigen PSI, Switzerland}

\author{P. Král}
\affiliation{PSI Center for Neutron and Muon Sciences CNM, 5232 Villigen PSI, Switzerland}

\author{I.~Maetsu}
\affiliation{Department of Engineering and Applied Sciences, Sophia University, 7-1 Kioi-cho, Chiyoda-ku, Tokyo 102-8554, Japan}

\author{H.~Gopakumar}
\affiliation{PSI Center for Neutron and Muon Sciences CNM, 5232 Villigen PSI, Switzerland}

\author{M.~M{\"u}ller}
\affiliation{PSI Center for Scientific Computing, Theory and Data, 5232 Villigen PSI, Switzerland}

\author{R.~Sarkar}
\affiliation{Institute for Solid State and Materials Physics, Technische Universität Dresden, D-01069 Dresden, Germany}

\author{V.~Grinenko}
\affiliation{Tsung-Dao Lee Institute and School of Physics and Astronomy, Shanghai Jiao Tong University, Shanghai 201210, China}

\author{G.~Simutis}
\affiliation{PSI Center for Neutron and Muon Sciences CNM, 5232 Villigen PSI, Switzerland}

\author{T.~Shiroka}
\affiliation{PSI Center for Neutron and Muon Sciences CNM, 5232 Villigen PSI, Switzerland}

\author{R.~Khasanov}
\affiliation{PSI Center for Neutron and Muon Sciences CNM, 5232 Villigen PSI, Switzerland}

\author{M.~Janoschek}
\affiliation{PSI Center for Neutron and Muon Sciences CNM, 5232 Villigen PSI, Switzerland}
\affiliation{Physik-Institut, Universität Zurich, Winterthurerstrasse 190, CH-8057 Zurich, Switzerland.}

\author{J.M.~Tranquada}
\affiliation{Condensed Matter Physics and Materials Science Division, Brookhaven National Laboratory, Upton, New York 11973, USA}

\author{H.H.~Klauss}
\affiliation{Institute for Solid State and Materials Physics, Technische Universität Dresden, D-01069 Dresden, Germany}

\author{T.~Adachi}
\affiliation{Department of Engineering and Applied Sciences, Sophia University, 7-1 Kioi-cho, Chiyoda-ku, Tokyo 102-8554, Japan}
    
\author{H.~Luetkens}
\email{hubertus.luetkens@psi.ch}
\affiliation{PSI Center for Neutron and Muon Sciences CNM, 5232 Villigen PSI, Switzerland}
	
\author{Z.~Guguchia}
\email{zurab.guguchia@psi.ch}
\affiliation{PSI Center for Neutron and Muon Sciences CNM, 5232 Villigen PSI, Switzerland}
	
\date{\today}
	
\begin{abstract}
The cuprate superconductor La$_{2-x}$Ba$_x$CuO$_4$ (LBCO) near $x=0.125$ is a striking example of intertwined electronic orders, where 3D superconductivity is anomalously suppressed, allowing spin and charge stripe order to develop, in a manner consistent with the emergence of a pair-density-wave (PDW) state. 
Understanding this interplay remains a key challenge in cuprates, highlighting the necessity of external tuning for deeper insight. While in-plane (within the CuO plane) uniaxial stress enhances superconductivity and suppresses stripe order, the effects of $c$-axis compression (perpendicular to the CuO plane) remains largely unexplored. Here, we use muon spin rotation ($\mu$SR) and AC susceptibility with an in situ piezoelectric stress device to investigate the spin-stripe order and superconductivity in LBCO-0.115 under $c$-axis compression. The measurements reveal a gradual suppression of the superconducting transition temperature ($T_{\rm c}$) with increasing $c$-axis stress, in stark contrast to the strong enhancement observed under in-plane stress. We further show that while in-plane stress rapidly reduces both the magnetic volume fraction ($V_{\rm m}$) and the spin-stripe ordering temperature ($T_{\rm so}$), $c$-axis compression has no effect, with $V_{\rm m}$ and $T_{\rm so}$ exhibiting an almost unchanged behavior up to the highest applied stress of 0.21\,GPa. These findings demonstrate a strong anisotropy in stress response, underscoring the critical role of crystallographic anisotropy in governing competing electronic phases in LBCO.	
\end{abstract}
\maketitle
	
\begin{figure*}[t!]
\centering
\includegraphics[width=1.0\linewidth]{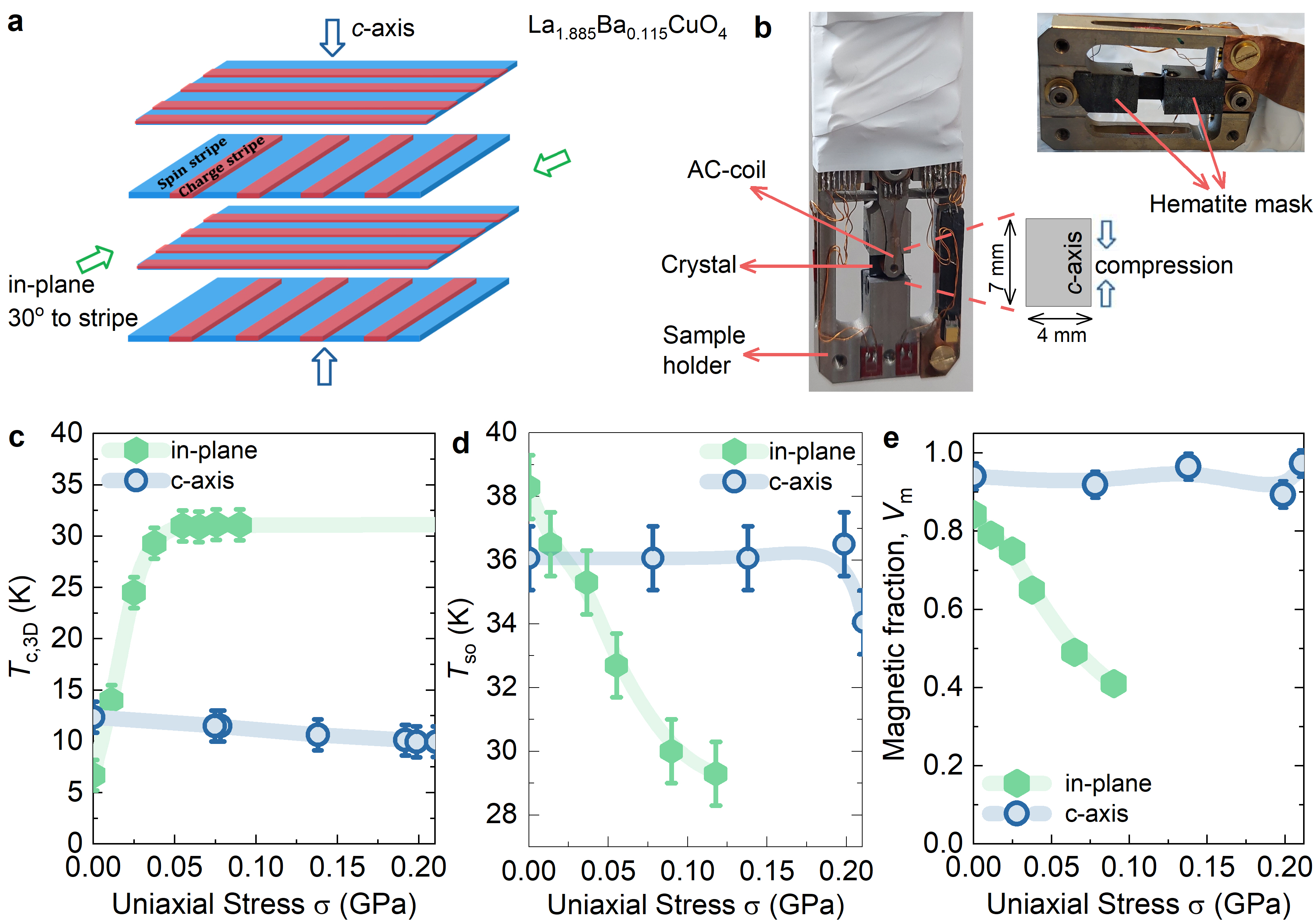}
\vspace{-0.4cm}
\caption{\textbf{a} Schematic representation of stripe order stacking along the $c$-axis in LBCO, showing charge and spin stripe modulation within  CuO$_2$ layers and their alternating 90$\degree$ rotation between adjacent layers. The blue and green arrows indicate the in-plane (at an angle of 30$^\circ$ to the stripe or Cu-O bond direction) and c-axis compressive stress, respectively. \textbf{b} The uniaxial stress sample holder used for the $\mu$SR experiments. Backside view of the sample showing the ACS coil and the Cernox temperature sensors (left panel). View from the direction of the incoming muon beam (right panel). Hematite pieces masking the holder frame exposed to the muon beam. \textbf{c}-\textbf{e} $c$-axis (hollow symbols) and in-plane (solid symbols) uniaxial stress dependence of \textbf{c} 3D SC transition temperatures $T_{\rm c,3D}$ derived from AC susceptibility measurements shown in Fig.~\ref{Fig2}\,a, \textbf{d} spin-stripe ordering temperatures ($T_{\rm so}$) derived from wTF-$\mu$SR experiments shown in Fig.~\ref{Fig3}\,a, and \textbf{e} the magnetically ordered fraction $V_{\rm m}$ at base-$T\simeq0.7$~K derived from ZF-$\mu$SR experiments shown in Fig.~\ref{Fig3}\,c, for La$_{1.885}$Ba$_{0.115}$CuO$_4$. The results for the in-plane strain are taken from Ref.~\cite{Guguchia097005}}
\label{Fig1}
\end{figure*}

Strongly correlated materials, particularly those hosting unconventional superconductivity, exhibit complex and intricate phase diagrams, where multiple broken-symmetry phases often emerge in response to external parameters such as doping, pressure, stress, impurities and magnetic fields~\cite{Fradkin457, Keimer179, Robinson126501,Guguchia097005,PhysRevLett.119.087002}. Among them, La$_{2-x}$Ba$_x$CuO$_4$ (LBCO) stands out as a compelling example harbouring a rich phase diagram of intertwined electronic orders~\cite{Keimer179}. Near a hole doping level of $x=0.125$, LBCO undergoes a dramatic suppression of the bulk three-dimensional (3D) superconducting (SC) transition temperature ($T_{\rm c, 3D}$)~\cite{Moodenbaugh4596}, coinciding with the emergence of static charge~\cite{Sears115125, Hu064504, Abbamonte155} and spin-stripe order~\cite{Tranquada561, Fujita104517, Hucker104506, Corboz046402}. This suppression is accompanied by a structural transition from the low-temperature orthorhombic (LTO) to the low-temperature tetragonal (LTT) phase, which introduces a lattice distortion, discussed as crucial for stabilizing the stripe order by anchoring it to the lattice potential~\cite{Hu064504, Axe2751, Axe271}. Within each of the CuO$_2$ layers, stripes align preferentially along a specific Cu-O bond direction, undergoing a 90$\degree$ rotation between adjacent layers, as illustrated in Fig.~\ref{Fig1}\,a. 

The strong suppression of 3D superconductivity at $x=0.125$ suggests an inherent incompatibility between stripe order and long-range SC coherence. However, experiments reveal a more nuanced picture, showing the coexistence of an unconventional 2D SC state with spin-stripe order~\cite{Tranquada174529, Li067001}. This phenomenon is interpreted in terms of a pair-density-wave (PDW) state, where SC pairing occurs along quasi-1D charge stripes with the phase of the pair wave function  shifting by $\pi$ from one charge stripe to the next~\cite{Agterberg231, Himeda117001, lee2023pair}. In adjacent CuO$_2$ layers, PDW states on orthogonal charge stripes frustrate the interlayer Josephson coupling, suppressing uniform 3D SC while preserving 2D SC correlations~\cite{Agterberg231, Berg127003}. This highlights the intricate electronic interactions in cuprates.

The major factors governing the interplay between spin-stripe order, PDW, and bulk superconductivity remain a subject of debate, with significant implications for understanding the mechanisms behind high-$T_{\rm c}$ superconductivity ~\cite{Emery8814, Dahm217, Huang1161,Tranquada184510, Kivelson1201, Vojta699, Parker2010, Choi2307515, Klauss4590}. Because these competing orders are delicately balanced, even small external perturbations can act as effective tuning parameters, selectively stabilizing different emergent phases. Among various tuning methods, hydrostatic pressure has been widely explored in LBCO. However, despite exceeding the critical pressure threshold where the long-range structural anisotropy (LTT phase - believed necessary for pinning charge stripes) is eliminated, the effect on $T_{\rm c}$ remains surprisingly modest~\cite{Hucker057004, Guguchia093005}. In contrast, uniaxial stress has proven to be a far more effective tool in manipulating the competing orders~\cite{Guguchia097005, Guguchiae2303423120, Thomarat271, BoyleL022004, Choi207002, Wang1795, Kim1040, Kamminga144506,simutis2022single,kuspert2024engineering}. Specifically, in-plane compression along the SC CuO$_2$ planes in LBCO ($x=0.115, 0.135$) has been shown to dramatically enhance $T_{\rm c}$ while simultaneously suppressing the magnetic volume fraction ($V_{\rm m}$), which is inversely correlated with 3D SC coherence~\cite{Guguchia097005, Guguchiae2303423120}. Additionally, in-plane (at an angle of 30$^\circ$ to the Cu-O bond direction) uniaxial stress of approximately $\sim 0.1$~GPa destabilizes the long-range LTT phase — thought to reinforce charge stripe order—while still preserving the onset temperature of spin-stripe order ($T_{\rm so}$), which intriguingly coincides with $T_{\rm c}$.

\begin{figure*} [t!]
\centering
\includegraphics[width=1.0\linewidth]{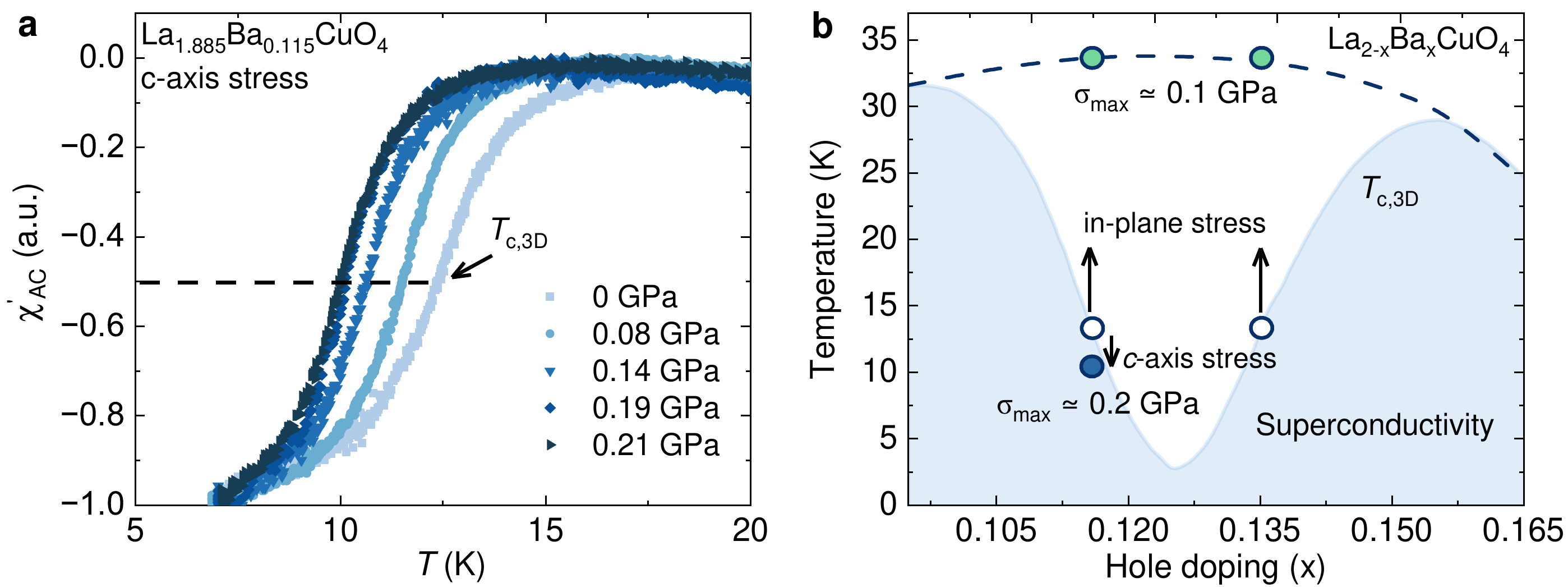}
\vspace{-0.5cm}
\caption{\textbf{a}~The temperature dependence of the AC diamagnetic susceptibility ($\chi'_{\rm AC}$) for La$_{1.885}$Ba$_0.115$CuO$_4$ under varying $c$-axis compressive stress. The arrow indicates the bulk 3D SC transition temperature ($T_{\rm c, 3D}$), defined as the point where $\chi'_{\rm AC} = -0.5$. \textbf{b}~Schematic temperature-doping ($x$) phase diagram of LBCO near $x=0.125$, illustrating the contrasting effects of uniaxial stress on superconductivity. The suppression of $T_{\rm c, 3D}$ under $c$-axis compression is shown for LBCO-0.115, while the pronounced enhancement of $T_{\rm c, 3D}$ under in-plane stress is depicted for both LBCO-0.115 and LBCO-0.135. The dashed line represents the anticipated SC phase boundary under in-plane stress across a broader doping range near $x=0.125$.}
\label{Fig2}
\end{figure*}

While the effects of in-plane stress have been studied extensively, the influence of uniaxial stress along the $c$-axis on spin-stripe order and superconductivity remains unexplored. Understanding this effect could provide crucial insights into the role of crystallographic anisotropy in governing electronic interactions in cuprates. In this study, we investigate the response of an LBCO-0.115 crystal~\cite{Adachi144524} to compressive stress applied along the $c$-axis, perpendicular to the SC CuO$_2$ layers, using an in situ piezoelectrically controlled stress device (Fig.~\ref{Fig1}\,b)~\cite{Hicks2018, ghosh2020piezoelectric, Grinenko2021}. To probe the microscopic response of spin-stripe order, we employ muon spin rotation ($\mu$SR) spectroscopy, while AC susceptibility measurements track the SC transitions. The summary of our results is shown in Fig.~\ref{Fig1}\,c-e, while the details of the AC susceptibility and $\mu$SR results are provided in Fig.~\ref{Fig2} and Fig.~\ref{Fig3} below, respectively. Unlike in-plane stress, which strongly enhances $T_{\rm c}$, compressive stress along the $c$-axis results in a modest suppression of $T_{\rm c}$ (Fig.~\ref{Fig1}\,c). Additionally, while in-plane stress leads to a rapid decrease in the onset of spin stripe order, $T_{\rm so}$ and the magnetic volume fraction ($V_{\rm m}$), both remain nearly constant up to the highest applied $c$-axis compression ($\sim 0.2$~GPa) (Fig.~\ref{Fig1}\,d and e). Our findings therefore reveal a pronounced anisotropy in how stripe-order and SC properties respond to $c$-axis versus in-plane uniaxial compression (Fig.~\ref{Fig1}\,c-e).


To investigate the impact of $c$-axis compressive stress on superconductivity in LBCO-0.115, in situ AC susceptibility measurements were performed either immediately before or after each $\mu$SR experiment at each applied stress. The excitation field was applied primarily perpendicular to the $c$-axis, to ensure that the shielding currents flowed perpendicular to the CuO$_2$ layers, which makes the measurement sensitive to the onset of three-dimensional (3D) superconductivity~\cite{Tranquada174529}. The results are presented in Fig.~\ref{Fig2}\,a. At zero applied stress, the diamagnetic response corresponds to measurements taken under the zero-force condition of the stress apparatus. To quantify the changes in $T_{\rm c}$, we define the midpoint of the transition as an indicator of 3D SC order ($T_{\rm c,3D}$), as marked in Fig.~\ref{Fig2}. The strongest diamagnetic signal is taken to represent the full SC volume fraction. Notably, as the compressive stress increases, a modest suppression of $T_{\rm c,3D}$ is observed, decreasing from 12.4\,K to 9.9\,K, with a linear suppression rate of 11.9\,K/GPa. As shown in Fig.~\ref{Fig2}\,b, this suppression occurs from an already reduced $T_{\rm c}$ at ambient pressure, further decreasing under the maximum applied $c$-axis compressive stress ($\sigma_{\rm max}\approx0.2$~GPa). In contrast, for both LBCO-0.115 and LBCO-0.135, an applied in-plane stress of $\sigma_{\rm max}\approx0.1$~GPa (only half the maximally applied $c$-axis stress) is sufficient to rapidly enhance $T_{\rm c,3D}$ to its optimal SC temperature observed in LBCO~\cite{Guguchia097005,Guguchiae2303423120}. This striking contrast underscores the strong anisotropic stress response of $T_{\rm c,3D}$.

\begin{figure*}[t!]
\centering
\includegraphics[width=1.0\linewidth]{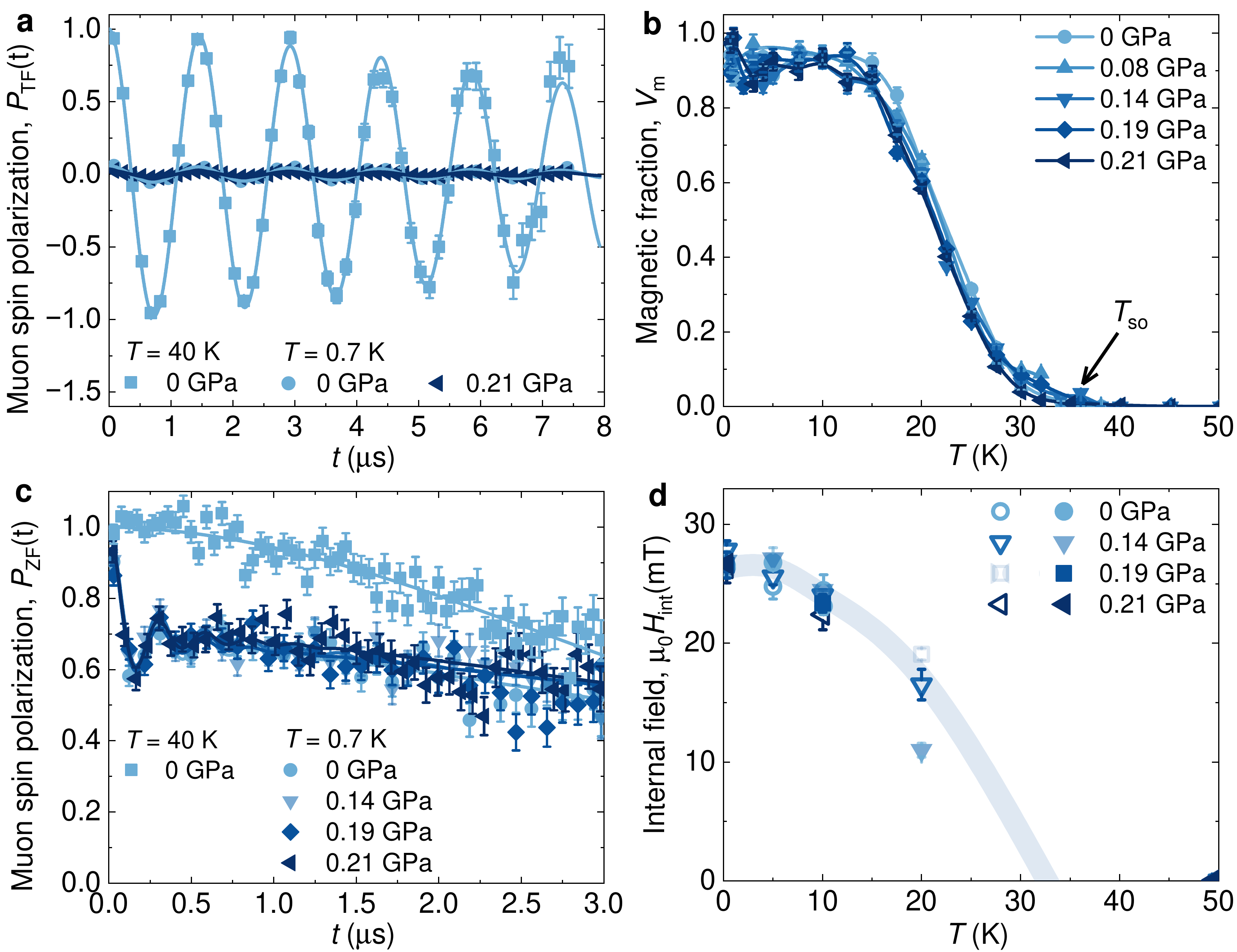}
\vspace{-0.5cm}
\caption{\textbf{a} Weak-TF $\mu$SR spectra of La$_{1.885}$Ba$_{0.115}$CuO$_4$ at 40 K under zero stress and at base-$T\simeq0.7$~K under varying $c$-axis compressive stress. \textbf{b} Temperature dependence of the magnetically ordered fraction ($V_{\rm m}$) under different stresses, extracted from Weak-TF $\mu$SR data in \textbf{a}. \textbf{c} Zero-field $\mu$SR spectra at 50 K under zero stress and $T\simeq0.7$~K under different $c$-axis stresses. \textbf{d} Temperature dependence of internal fields under various stresses, derived from ZF-$\mu$SR data in \textbf{c}. Solid and empty symbols indicate the values extracted for forward-backward and left-right detector sets (see Methods section), respectively.} 
\label{Fig3}
\end{figure*}

To uncover the evolution of spin-stripe order under compressive stress, we performed a combination of weak transverse-field (wTF) and zero-field (ZF) $\mu$SR measurements. The $\mu$SR technique provides a unique capability to probe the local magnetic order parameter and the magnetically ordered volume fraction independently, making it an invaluable tool for distinguishing between phase separation and homogeneous magnetic order. This is particularly crucial in cuprates, where competing electronic phases often coexist, leading to microscopic inhomogeneity \cite{Guguchia097005,Guguchiae2303423120}.
In a typical $\mu$SR experiment, spin-polarized muons are implanted into the sample, where their spins precess around the local internal magnetic field. The muons subsequently decay into positrons, which are emitted preferentially along the initial muon spin direction. By monitoring the time evolution of positron decay, one can extract information about the muon spin polarization $P(t)$. (Further details on the methodology can be found in the methods section.)


Figure~\ref{Fig3}\,a presents a selection of muon-spin polarization $P_{\rm TF} (t)$ curves measured under a weak transverse field (wTF) of 5\,mT, as a function of increasing $c$-axis compressive stress. In wTF-$\mu$SR, muons implanted in a nonmagnetic environment exhibit long-lived precessional oscillations with maximum amplitude. The initial polarization $P_{\rm TF} (0)$ in the paramagnetic state thus serves as a direct measure of the paramagnetic volume fraction. In contrast, when magnetic order is present, the implanted muons experience a broad distribution of internal fields, leading to rapid dephasing and a reduction in the oscillation amplitude. This allows for a precise determination of the magnetically ordered volume fraction, which can be estimated using the relation $V_{\rm m}=1-P_{\rm TF}(0)$. At 40\,K and zero applied stress, $P_{\rm TF}(0)$ remains close to 1, confirming the absence of magnetic order. However, at 0.7\,K, the value decreases significantly to $P_{\rm TF}(0)\approx 0.1$ and remains unchanged up to the maximum applied stress, indicating the development of spin-stripe order in the majority of the sample volume. The temperature dependence of the extracted magnetic volume fraction ($V_{\rm m}$) under varying $c$-axis compressive stress is shown in Fig.~\ref{Fig3}\,b. The onset of spin stripe ordering $T_{\rm so}$ and $V_{\rm m}$ remains almost unchanged, independent of the applied pressure up to $\sim 0.21$~GPa. These results suggest that both $T_{\rm so}$ and $V_{\rm m}$ are largely insensitive to $c$-axis stress, highlighting the robustness of stripe order in LBCO-0.115 despite $c$-axis uniaxial perturbations. Notably, at the highest applied stress, $T_{\rm c,3D}$ remains almost three times smaller than $T_{\rm so}$.

Figure~\ref{Fig3}\,c illustrates the evolution of zero-field (ZF) muon spin polarization $P_{\rm ZF}(t)$ in LBCO-0.115 under increasing $c$-axis compressive stress. In ZF-$\mu$SR measurements, the implanted muon spins precess solely in response to the internal magnetic field inside a sample. The observed signal reflects the collective behavior of muon spins, averaging over their spatial distribution relative to local variations in the internal field. At high temperature (50\,K) and zero applied stress, $P_{\rm ZF}(t)$ exhibits a weakly decaying Gaussian-like behavior, characteristic of the paramagnetic regime. In this state, muon depolarization is primarily governed by static nuclear dipole moments, as the electronic moments fluctuate too rapidly to significantly influence the implanted muons. In contrast, at $T\simeq0.7$~K ($T\ll T_{\rm so}$), $P_{\rm ZF}(t)$ develops rapidly damped oscillations at early times ($t<1\mu$s), a hallmark of long-range spin-stripe order~\cite{Guguchia214511, Klauss4590}. As shown in Fig.~\ref{Fig3}\,c, these oscillations persist even as the compressive stress increases, with their amplitude remaining largely unchanged up to the highest applied stress. In single-crystal ZF-$\mu$SR measurements, the oscillation amplitude is influenced by both the magnetically ordered volume fraction ($V_{\rm m}$) and the angle between the muon spin and the internal magnetic field ($\mu_0H_{\rm int}$). Assuming that the stress does not alter the direction of the internal field at the muon stopping site, the observed insensitivity of the oscillation amplitude to applied stress is likely due to the unaltered behavior of $V_{\rm m}$, consistent with findings from wTF-$\mu$SR experiments. The characteristic internal field ($\mu_0H_{\rm int}$) at the muon stopping site, extracted from the oscillation frequency (as detailed in methods section), also remains mostly unchanged under increasing stress as shown in Fig.~\ref{Fig3}\,d. This indicates that the spin-stripe order maintains its long-range nature even at the highest applied $c$-axis compressive stress, further underscoring its robustness in LBCO-0.115 against $c$-axis uniaxial stress.

\begin{figure} [t!]
\centering
\includegraphics[width=1.0\linewidth]{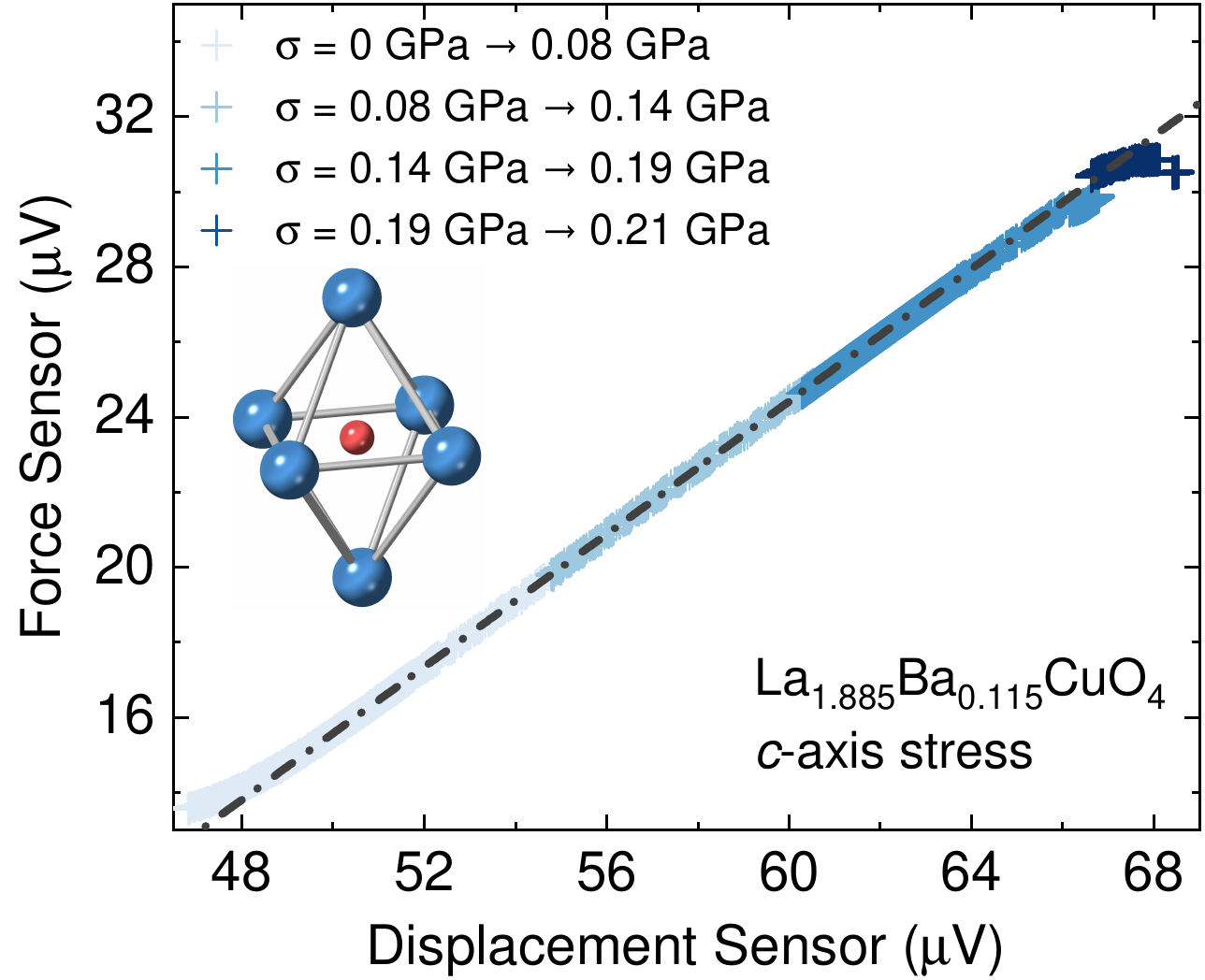}
\vspace{-0.5cm}
\caption{Strain-stress curve for LBCO-0.115. The force sensor output is proportional to strain, while the displacement sensor reading is proportional to stress. The dashed lines represent linear fits to the data across the entire applied stress region. The inset shows the distorted CuO octahedron in the LTT structure.}
\label{Fig4}
\end{figure}

Two important aspects can be highlighted from our results. First, they demonstrate a significant difference between the effects of c-axis and in-plane stress. Additionally, $c$-axis stress leads to a modest reduction in $T_{\rm c,3D}$, while the spin-stripe order remains completely unaffected. The stability of spin-stripe order is supported by our observation of a perfectly linear stress ($c$-axis)–strain (force-displacement) response (as shown in Fig.~\ref{Fig4}), providing indirect evidence that the LBCO-0.115 crystal structure retains its LTT symmetry. In contrast, a structural transition would typically manifest as a change in the slope of the stress-strain dependence~\cite{Guguchia097005}. One possible explanation for the absence of a $c$-axis stress effect on stripe order is the following: as shown recently using X-ray diffraction \cite{Sears115125} on LBCO-0.125, the screening of the stripes is correlated between the layers through the displacement of La ions and the apical oxygen atoms. These correlated displacements play a crucial role in stabilizing the stripe order across adjacent layers. Since $c$-axis strain does not break the existing symmetry of the crystal structure, it is unlikely to disrupt this interlayer coupling mechanism. In contrast to in-plane strain, which can directly affect the charge and spin stripe periodicity by modifying the CuO$_{2}$ plane geometry, $c$-axis stress primarily alters interlayer spacing without significantly perturbing the essential electronic interactions that underpin stripe order. As a result, the stripe order remains robust despite the applied stress along the $c$-axis. This interpretation aligns with the observed resilience of stripe order under $c$-axis stress and highlights the role of interlayer interactions in governing the stability of charge and spin stripes in LBCO.

An important question arises: if the stripe order remains unchanged, why does $T_{\rm c,3D}$ decrease under stress? To gain insight, we recall that the strong suppression of 3D superconductivity (as shown in Fig.~\ref{Fig2}\,b) in LBCO near $x$=0.125 is typically explained ~\cite{Fradkin457, Keimer179, Agterberg231} by the fact that the anti-phase 2D SC states (PDW state) residing within the orthogonal stripes in adjacent layers disrupt interlayer phase coherence. This, in turn, frustrates the interlayer Josephson coupling and naturally explains the suppression of 3D SC order at ambient pressure. The gradual suppression of $T_{\rm c,3D}$ up to highest applied pressure $\sim0.21$~GPa indicates that $c$-axis stress further disrupts 3D SC coherence. One plausible explanation is that $c$-axis stress enhances frustration of the inter-layer Josephson coupling. This can be understood by considering the coexistence of the PDW and 3D uniform $d$-wave SC regions. In adjacent layers, the 2D $d$-wave SC patches, whose projections (along $c$) overlap, mediate finite inter-layer coupling. However, if such patches remain sparse, PDW order dominates intra-layer physics, leading to frustrated inter-layer couplings. In-plane uniaxial stress has been shown to suppress stripe order and enhance bulk 3D superconductivity until $\sigma_{\rm cr}\simeq0.05$~GPa, thereafter $T_{\rm so}$ and $T_{\rm c,3D}$ level off and coincide with each other~\cite{Guguchia097005, Guguchiae2303423120}. It has been well understood that in-plane stress can suppress stripe order by inducing significant distortions in the crystal structure, ultimately breaking the low-temperature tetragonal (LTT) symmetry at $\sigma_{\rm cr}\simeq0.05$~GPa. Subsequently, this structural modification facilitates the formation of uniform SC patches in adjacent layers, enhancing inter-layer Josephson coupling and driving the system toward bulk 3D superconductivity. Conversely, the steady suppression of $T_{\rm c,3D}$ under $c$-axis stress cannot be attributed to an enhancement of spin-stripe order as $V_{\rm m}$ remains nearly unchanged. Moreover, $c$-axis stress does not alter the crystal symmetry (as shown in Fig.~\ref{Fig4}). However, stress likely induces subtle local lattice distortions that modify the spatial distribution of d-wave superconducting (SC) patches. Expanding on this, the charge density within the two-dimensional d-wave patches differs from both the layer-averaged charge density and that of the stripy regions. Consequently, adjacent patches likely experience Coulomb repulsion. Stress applied along the $c$-axis enhances this effect by reducing the interlayer distance, which could, in turn, favor an anticorrelated arrangement of patches in neighboring layers, thereby decreasing their overlap. This would further enhance frustration in the interlayer Josephson coupling.

Although our measurements do not directly capture the onset of two-dimensional (2D) SC order ($T_{\rm c,2D}$), it can be reasonably approximated based on the well-established fact that, under ambient pressure, $T_{\rm c,2D} \approx T_{\rm so}$~\cite{Tranquada174529, Li067001, Guguchia214511}. Consequently, the significant energy scale separation between $T_{\rm c,3D}$ and $T_{\rm so}$ under $c$-axis stress strongly suggests that 2D SC order (PDW state) persists despite the suppression of 3D superconductivity. This reminds us of previous studies on LBCO ($x\approx0.095, 0.125$)~\cite{Lieaav7686, Wen134513, Stegen064509,PhysRevLett.95.157001} and La$_{2-x}$Sr$_x$CuO$_4$ ($x\approx0.1$)~\cite{Schafgans157002}, where a strong $c$-axis magnetic field was found to completely suppress 3D superconductivity while leaving 2D SC correlations intact. Both $c$-axis magnetic field and even though the effect is small as we show here $c$-axis stress disrupt 3D SC coherence; however, a key distinction is that a strong $c$-axis field has been reported to enhance spin-stripe order~\cite{Lake299, Wen134513}. In contrast, our results for LBCO-0.115 show no such effect, as the stripe ordered fraction remains largely unaffected by $c$-axis stress, coexisting with 3D and 2D SC orders.

In conclusion, we investigated the evolution of spin-stripe order and superconductivity in LBCO-0.115 under $c$-axis compressive stress using muon spin rotation and AC susceptibility measurements. Our results reveal that while stress induces a modest suppression of 3D superconductivity, spin-stripe order remains largely unaffected. The persistence of a significant energy scale separation between $T_{\rm c,3D}$ and $T_{\rm so}$ throughout the applied $c$-axis stress regime suggests that 2D SC correlations continue to coexist with the spin-stripe order in this system. The contrasting response of stripe and SC correlations to in-plane and $c$-axis stress underscores the profound role of crystallographic anisotropy in controlling phase competition in LBCO. Our findings emphasize the need for a comprehensive theoretical framework to understand the stress-driven interplay between 3D superconductivity and spin order in the stripe phase of cuprates, offering potential insights into high-$T_{\rm c}$ superconductivity. They also pose critical questions regarding the precise impact of $c$-axis stress on crystal structure and charge-stripe order, emphasizing the need for further experimental exploration. 

As an outlook and future direction, an intriguing avenue for exploration is the behavior of interlayer plasmons \cite{Laplace03052016} and their evolution in response to a potential reorganization of d-wave superconducting patches. Given the interplay between charge density variations and stress-induced changes in interlayer spacing, the spatial arrangement of these patches could significantly impact plasmonic excitations. In particular, an anticorrelated positioning of patches in neighboring layers may disrupt the coherence of interlayer Josephson coupling, leading to distinct plasmonic signatures. Understanding this evolution could offer valuable insights into the role of electronic inhomogeneity in shaping collective charge dynamics in layered superconductors.

\section{Acknowledgments}~
The ${\mu}$SR experiments were carried out at the Swiss Muon Source (S${\mu}$S) Paul Scherrer Institute, Villigen, Switzerland. Z.G. acknowledges support from the Swiss National Science Foundation (SNSF) through SNSF Starting Grant (No. TMSGI2${\_}$211750). Work at Brookhaven is is supported by the Office of Basic Energy Sciences, Materials Sciences and Engineering Division, U.S. Department of Energy under Contract No.\ DE-SC0012704.\\

\section{Author contributions}~
Z.G. conceived and supervised the project. Sample growth: I.M., T.A..
Preparation for strain experiments: S.S.I., V.S., H.G., H.L., and Z.G.. $\mu$SR and AC susceptibility experiments, the corresponding analysis and discussions: S.S.I., V.S., J.N.G., O.G., P.K., H.G., M.M., R.S., V.G., G.S., T.S., R.K., M.J., J.T., H.H.K., H.L., and Z.G.. Figure development and writing of the paper: S.S.I. and Z.G. All authors discussed the results, interpretation, and conclusion.\\

\section{Methods}

\subsection{Principles of the \(\mu\)SR Technique}

\begin{figure*} [t!]
\centering
\includegraphics[width=1.0\linewidth]{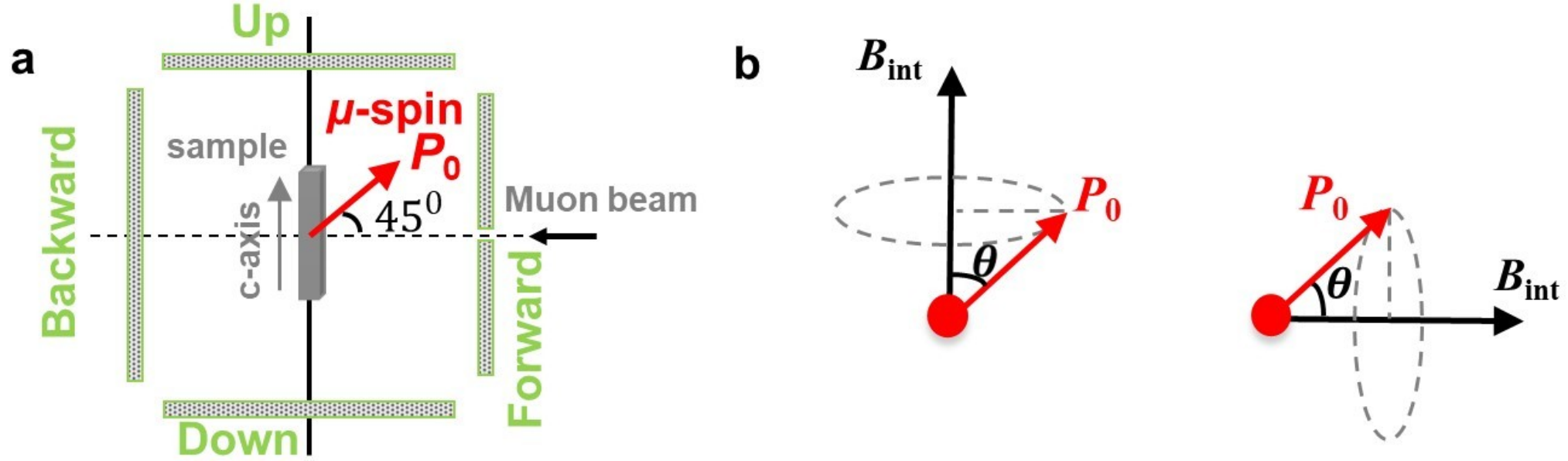}
\vspace{-0.1cm}
\caption{\textbf{a} Schematic representation of the experimental setup, where the muon spin is oriented at a 45$^\circ$ angle relative to the muon beam. The sample is surrounded by four positron detectors: Forward (F), Backward (B), Left (L), and Right (R). \textbf{b} Illustration of muon spin precession around the internal magnetic field in two scenarios:  (Left) The field is perpendicular to the \( c \)-axis, directed toward the L-detector, with \( \theta \) representing the angle between the magnetic field and the initial muon spin polarization at \( t = 0 \).  
(Right) The field is aligned parallel to the \( c \)-axis and oriented toward the F-detector. }
\label{FigS1}
\end{figure*}

In a typical $\mu$SR experiment (see Fig.~\ref{FigS1}\,a-b), a high-intensity beam of 100\% spin-polarized positive muons is directed into the sample, where the muons thermalize at interstitial lattice sites and act as highly sensitive local magnetic probes. The muons, carrying a momentum of $p_{\mu} = 29$~MeV/c, experience the local magnetic field $B_{\mu}$, causing their spins to precess at the Larmor frequency $\omega_{\mu} = 2\pi \nu_{\mu} = \gamma_{\mu} B_{\mu}$, where $\gamma_{\mu} / (2\pi) = 135.5$~MHz/T is the muon gyromagnetic ratio.

The muon-spin relaxation ($\mu$SR) experiments under zero field (ZF) and weak transverse field (wTF), the latter being applied parpenicular to the initial muon-spin polarization, were performed at the Dolly spectrometer
($\pi$E1 beamline) at the Paul Scherrer Institute, Villigen, Switzerland. 
In this technique, the implanted muons decay with a mean lifetime of $\tau_{\mu} = 2.2~\mu$s, emitting positrons preferentially along the muon spin direction. A set of detectors surrounding the sample records the arrival of muons and the subsequent emission of positrons. The detection process begins when a muon enters the sample, initiating an electronic clock. This clock stops when the decay positron is detected in one of the positron detectors, and the time interval is stored in a histogram. This process is repeated for millions of muon decay events, creating a time-resolved positron count histogram.

In the Dolly instrument, the sample is surrounded by four positron detectors—Forward, Backward, Left, and Right—positioned relative to the muon beam direction. The recorded positron counts in these detectors, $N_{\alpha}(t)$ (where $\alpha = F, B, L, R $), follow an exponential decay due to the muon's finite lifetime. This count distribution also includes a time-dependent polarization function $P(t)$, which encodes the information about the local magnetic environment. The equation governing this decay is given by:

\begin{equation}
N_{\alpha}(t) = N_0 e^{-t/\tau_{\mu}} \left[ 1 + A_0 P(t) n_{\alpha} \right] + N_{\text{bg}}
\end{equation}

where $N_0$ is proportional to the number of recorded events, $A_0$ is the initial asymmetry factor dependent on detector geometry and positron scattering, and $N_{\text{bg}}$ represents the background contribution from uncorrelated events. The initial asymmetry $A_0$ typically ranges between 0.2 and 0.3.

Since positrons are preferentially emitted along the muon spin direction, the signals recorded by the Forward and Backward detectors exhibit oscillations corresponding to the muon precession frequency. To remove the exponential decay component associated with the muon's finite lifetime, a reduced asymmetry function $A(t)$ is introduced:

\begin{equation}
A(t) = \frac{N_{F,L}(t) - N_{B,R}(t)}{N_{F,L}(t) + N_{B,R}(t)} = A_0 P(t)
\end{equation}

where $N_{F,L}(t)$ and $N_{B,R}(t)$ denote the positron counts recorded in the Forward/Left and Backward/Right detectors, respectively. The asymmetry function $A(t)$ and the polarization function $P(t)$ provide direct insight into the static and dynamic properties of the local magnetic environment at the muon site. These functions serve as a powerful tool for investigating the spatial distribution of magnetic fields and their fluctuations in complex materials.

\subsection{Analysis of ZF-$\mu$SR data}  

The ZF-$\mu$SR signals shown in were analyzed over the entire temperature range by decomposing the asymmetry signal into contributions from both magnetic and nonmagnetic regions. The zero-field $\mu$SR polarization function is expressed as:  
\begin{equation}
\begin{split}
P_{\text{ZF}}(t) = & V_{\rm m} \left[ f_{\alpha} e^{-\lambda_T t} J_0(\gamma_{\mu} B_{\text{int}} t) + (1 - f_{\alpha}) e^{-\lambda_L t} \right] \\ + 
& (1 - V_{\rm m}) e^{-\lambda_{\text{nm}} t}.
\end{split}
\end{equation}

Here, $V_{\rm m}$ represents the magnetically ordered volume fraction, while $B_{\text{int}}$ denotes the maximum internal field associated with the Overhauser distribution. The depolarization rates $\lambda_T$ and $\lambda_L$ characterize the transverse and longitudinal relaxation of the magnetic regions, respectively. The parameters $f_{\alpha}$ and $(1 - f_{\alpha})$ correspond to the fractions of the oscillating and non-oscillating components of the magnetic $\mu$SR signal. The function $J_0$ is the zeroth-order Bessel function of the first kind, which describes the incommensurate nature of the spin-density wave and accounts for broad internal field distributions spanning from zero to a maximum value. This behavior is commonly observed in cuprates exhibiting static spin-stripe order~\cite{Guguchia097005, Guguchia214511, Guguchiae2303423120, Klauss4590}. $\lambda_{\text{nm}}$ represents the relaxation rate of the nonmagnetic volume fraction, where spin-stripe order is absent. The complete analysis of $\mu$SR time spectra, including both zero-field (ZF) and transverse-field (TF) measurements, was performed using the open-source $\mu$SR data fitting software \textbf{musrfit}~\cite{Suter2012}.

\subsection{Analysis of Weak-TF \(\mu\)SR Data}

The weak-transverse-field (\(\mu\)SR) asymmetry spectra were analyzed using a model describing the time evolution of the muon-spin polarization:

\begin{equation}
P_{\text{TF}}(t) = P_{\text{TF}}(0) e^{-\lambda t} \cos(\omega t + \phi),
\end{equation}

where \( P_{\text{TF}}(t) \) represents the muon-spin polarization at time \( t \) after implantation, \( P_{\text{TF}}(0) \) denotes the initial polarization amplitude linked to the paramagnetic volume fraction, and \( \lambda \) is the depolarization rate associated with paramagnetic spin fluctuations and nuclear dipolar interactions. The frequency \( \omega \) corresponds to the Larmor precession, which depends on the applied transverse magnetic field, while \( \phi \) accounts for the phase offset.

For magnetically ordered samples, baseline asymmetry shifts were corrected by allowing \( P_{\text{TF}}(t) \) to be adjusted for each temperature. The magnetically ordered volume fraction at temperature \( T \) was determined using the relation:

\begin{equation}
V_{\rm m}(T) = 1 - P_{\text{TF}}(0,T).
\end{equation}

In the high-temperature paramagnetic phase (\( T > T_{\rm so} \)), where no long-range magnetic order exists, the initial polarization reaches its maximum value, \( P_{\text{TF}}(0,T > T_{\rm so}) = 1 \).

In weak-TF \(\mu\)SR measurements, the signal typically consists of long-lived oscillations from muons precessing in the applied external field, alongside strongly damped oscillations from muons in magnetically ordered regions, where they experience a broad internal field distribution due to the combined influence of applied and internal magnetic fields. ZF-\(\mu\)SR measurements reveal rapid depolarization on a timescale of approximately 0.5~\(\mu\)s. However, in TF-\(\mu\)SR data, variations in the angle between the applied and internal fields create a highly inhomogeneous field distribution in magnetically ordered regions, leading to significant muon-spin dephasing. This, combined with data binning, results in the damped signals from magnetic patches being effectively suppressed in the weak-TF \(\mu\)SR spectra.
	
\bibliography{LBCO_c-axis_stress}

\end{document}